\documentstyle[12pt,epsf]{article}

\newcommand{\beq}{\begin{equation}}
\newcommand{\eeq}{\end{equation}}

\begin{document}
\def\lag{\langle}
\def\rag{\rangle}
\def\today{\ }
\title{Multicanonical Simulation of the van Hemmen Spin
 Glass\thanks{To appear in the Proceedings of the Sixth Annual
Workshop on Recent Developments in Computer Simulation Studies in Condensed
Matter Physics, 22--26 Feb. 1993, Athens, Georgia.}}

\author{M.\ Katoot$^1$, U.\ Hansmann$^{2,3}$ and T.\ Celik$^{3,4}$}
\footnotetext[1]{{\em
 Department of Mathematics and Physical Sciences,\
 Embry-Riddle Aeronautical University,\ Daytona Beach,\ FL 32114,
 USA}}
\footnotetext[2]{{\em
 Department of Physics, The Florida State University, Tallahassee,
 FL 32306, USA}}
\footnotetext[3]{{\em
 Supercomputer Computations Research Institute Tallahassee, The
Florida
 State University, Tallahassee, FL 32306, USA}}
\footnotetext[4]{{\em On leave of absence from Department of
            Physics Engineering, Hacettepe University, Ankara,
	    Turkey}}
\maketitle
\pagestyle{empty}
\begin{abstract}
We studied the performance of the multicanonical algorithm by
simulating
the van Hemmen spin glass model and reproduced the exact results
for
this mean field model. Physical quantities such as energy density,
specific heat, susceptibility
and order parameters are evaluated at all temperatures.
\end{abstract}

One of the recent and promising developments in computational
physics is the
use of the multicanonical ensemble\cite{BN1} for numerical
simulations.
It was originally developed
for systems with first order phase transitions to avoid the
supercritical
slowing down \cite{BN2,BHN2,JBK}.
Another possible target of the method are systems with conflicting
constraints.  For low temperatures these systems split into many
thermodynamic states, separated by high tunneling barriers. Low
temperature
canonical simulations tend to get trapped in one of those states.
Multicanonical simulations overcome these barriers by connecting
back to high temperature states. First promising studies exist for
spin glasses \cite{BC1,BC2,BCH}, random ising model \cite{MP} and
proteins \cite{HO}. Here, we test the performance of the
multicanonical
algorithm  against exact results for this kind of systems, namely
the van Hemmen spin glass
model \cite{VH}.

In this paper, due to limited space, we will not review the
multicanonical ensemble.  Interested readers should refer to the
references \cite{BN1,BN2,BC1,BHN1}.

The van Hemmen spin glass model\cite{VH} is defined by the hamiltonian:
\begin{equation}
H = -\frac{J_0}{N} \sum_{i,j} S(i)S(j) - \sum_{i,j} J_{ij}S(i)S(j)
- h
\sum_i S(i),
\end{equation}
describing N Ising spins interacting with an external magnetic
field $h$
and with each other in pairs $(i,j)$. Via $J_0$  a direct
ferromagnetic
coupling is incorporated. The $J_{ij}$ contain the randomness,
\begin{equation}
J_{ij} = \frac{J}{N} [\xi_i \eta_j + \xi_j \eta_i ],
\end{equation}
where the $\xi_i$ and $\eta_j$ are independent and evenly
distributed random variables with mean zero and variance one. This
distribution of  $J_{ij}$ 's is shown \cite {BS} to model the RKKY
\cite{RKKY} interaction
in a real metallic spin glass:
symmetric and highly peaked at $J_{ij} = 0$. The $J_{ij}$ contain
$2N$
independent random variables and describe therefore a random-site
problem,
not a random-bond problem like the SK model.

The model has three order parameters,
\begin{equation}
m_N = N^{-1}  \sum_{i=1}^{N} S(i) ,~~~~~~~ q_{1N} = N^{-1}
\sum_{i=1}^{N} \xi_i
S(i) ,
{}~~~q_{2N} = N^{-1} \sum_{i=1}^{N} \eta_i S(i).
\end {equation}
Without a ferromagnetic interaction, i.e., for $J_0 = 0$ the
magnetization
vanishes  $m=0$, and the order parameters $q_{1N}$ and $q_{2N}$
are
 combined to give a more relevant order parameter $Q$:
\begin{equation}
Q = N^{-1} < {1\over 2} ( q_{1N} + q_{2N} ) > ,
\end{equation}
from which the thermodynamical quantities  can be obtained.
This model is exactly solvable without replicas and its main
features are
consistent with that of a spin glass model with randomness and
frustration,
except for the metastable state structure.  The system  actually
picks out
 a Mattis state \cite{Matt} and lacks  therefore a great
multiplicity of
metastable states which is considered integral to a true spin
glass.
We like to mention that the van Hemmen model
 is also very closely related to the Hopfield model
\cite{Hopf} which is widely used to model neural networks.

We performed multicanonical simulations of the van Hemmen model on
cluster of RISC workstations. Independent gaussian distributions
for $\xi_i$
and $\eta_i$ with mean value zero and variance one were created.
Simulations
with up to $N = 1000 $ spins were easily carried out.
Thermodynamical averages were evaluated
over
typically two million iterations, following $2\times 10^5$
iterations of
thermalization runs. With this extended statistic and the new
method
we could go far beyond what was done in earlier work\cite{HM}.
\begin{figure}[htb]
\vspace{7cm}
\includegraphics{mg10.ps}
\caption{Spin glass order parameter distribution.}
\label{fig:figure 1}
\end{figure}
Let us first concentrate on the pure spin glass case and set the
ferromagnetic coupling $J_0 = 0$. Fig.1 shows the distribution of
the
order parameter $Q$ at all temperatures for $N = 1000$ spins .
$T_{f}/J = 1$ is the bifurcation point below which temperature the
nonzero
ordering sets in and reaches its maximum value $1 / {\sqrt \pi}$ at
$T = 0$.
\begin{figure}[htb]
\vspace{5.9cm}
\includegraphics{spcht2.ps}
\caption{Specific heat vs. temperature for several lattice sizes.}
\label{fig:figure 2}
\end{figure}
There is no many-valley
structure like the one observed in the multicanonical simulation of
Edwards-Anderson model \cite{BC2}. The internal energy assumes the
values $0$ at $T = T_f$ and   $-1/ \pi$ at $T = 0$.
Fig.2 displays the specific heat vs. temperature
for several lattice sizes. The specific heat is linear at low
temperatures,
peaks at $0.8 < T < 1.0$ and vanishes for $T_f < T$. The linear
behavior
of the specific heat indicates the existence of many low lying
excited
states at low temperatures. Both our values for the internal energy
and specific heat reproduce the exact values with high precision.

Next we included the ferromagnetic coupling in the model, with
zero external field. The spin glass to ferromagnet transition is
supposed to
take place in the region $J_0 \sim J$.
We observed that the magnetization jumps and
the order respectively vanishes for $J_0$ approaching $J$. While at
$J_0 / J = 0.6$ the distribution of the order parameter was the
same as
depicted in Fig.1, for $J_0 / J = 0.9$ it assumed the value $Q =
0.34$
at about $T / J \sim 2$ and stayed constant all the way down to $T
= 0$.
The system  stays in the metastable spin glass phase, and
does not spontaneously jump to the ferromagnetic phase. Such
behavior was
observed elsewhere \cite{GWL}.
In Fig.3 we show the zero field susceptibility $\chi_{0} (T)$ vs.
temperature for several values
of the ferromagnetic coupling $J_0$. For $T < T_f$ the
susceptibility
develops a plateau, a feature indicating an equilibrium phenomena,
and also shared by the Sherrington-Kirkpatrick
model. The value of the susceptibility at $T = T_f$ agrees well
with the
analytic solution $\chi_{0} (T_f) = 1/(J-J_0)$. For $T_f < T$, we
observe
a pure Curie-Weiss behavior $\chi_{0} (T) = 1/(T-T_f)$.
Entering the region of the metastable spin glass phase mentioned
above,
around $J_0 / J = 0.6$, the plateau starts getting distorted.

\begin{figure}[htb]
\vspace{5.9cm}
\includegraphics{suscept2.ps}
\caption{Susceptibility vs. temperature for several
values of the ferromagnetic coupling.}
\label{fig:figure 3}
\end{figure}
The multicanonical
simulation of the van Hemmen model is quite feasible; the phases
are
clearly observed, main features like linear rising  specific heat,
the plateau in zero-field susceptibility are easily obtained.
Resulting
numerical values agree remarkably well with the predicted ones. It
is important to notice that these thermodynamic quantities were
obtained for all temperatures from one simulation.
In our simulations it is straightforward to probe the ground states
for
systems like the present one which includes
randomness and frustration and  a continuous energy spectrum.
The multicanonical algorithm reveals itself to be a reliable
instrument
for the simulation of systems with conflicting constraints and a
useful tool to study NP-complete systems.
\\
\\
\\
{\Large \bf Acknowledgments}

This research project was partially funded by the
Department of Energy under contract
DE-FC05-85ER2500 and the NATO Science Program.
T.C. was supported by TUBITAK of Turkey and U.H by the Deutsche
Forschungsgemeinschaft \hbox{under} contract H180411-1.

\end{document}